\documentclass[aps, reprint, superscriptaddress]{revtex4-1}
\usepackage{amssymb}
\usepackage{bm}
\usepackage{graphicx}
\usepackage{hyperref}
\newcommand{\mvec}[1]{\ensuremath{\mathbf{#1}}}

\newcommand{\Heff}{\ensuremath{\mvec{H}_\mathrm{eff}}}

\newcommand{\ttau}{\tilde{\tau}}
\newcommand{\tausd}{\tau_\mathrm{sd}}
\newcommand{\tausf}{\tau_\mathrm{sf}}
\newcommand{\dm}{\delta \mvec{m}}

\begin{document}
\title{Current-induced instability of domain walls in cylindrical nanowires}


\author{Weiwei Wang}
\email{wangweiwei1@nbu.edu.cn}
\affiliation{Department of Physics, Ningbo University, Ningbo 315211, China}
\author{Zhaoyang Zhang}
\affiliation{Department of Physics, Ningbo University, Ningbo 315211, China}
\author{Ryan A. Pepper}
\affiliation{Engineering and the Environment, University of Southampton,  SO17 1BJ, Southampton, United Kingdom}
\author{Congpu Mu}
\affiliation{School of Science, Yanshan University, Qinhuangdao 066004, China}
\author{Yan Zhou}
\affiliation{School of Science and Engineering, Chinese University of Hong Kong (Shenzhen), China}
\author{Hans Fangohr}
\affiliation{Engineering and the Environment, University of Southampton,  SO17 1BJ, Southampton, United Kingdom}
\affiliation{European XFEL GmbH, Holzkoppel 4, 22869, Schenefeld, Germany}


\begin{abstract}
We study the current-driven domain wall (DW) motion in
cylindrical nanowires using micromagnetic simulations by implementing the
Landau-Lifshitz-Gilbert equation with nonlocal spin-transfer torque in a finite difference
micromagnetic package. We find that in the presence of DW Gaussian wave packets (spin waves)  will be generated
when the charge current is applied to the system suddenly. And this effect is excluded when using
the local spin-transfer torque. The existence of spin waves emission indicates that transverse domain walls can not
move arbitrarily fast in cylindrical nanowires although they are free from the Walker limit.
We establish an upper velocity limit for the DW motion by analyzing the stability
of Gaussian wave packets using the local spin-transfer torque. Micromagnetic simulations
show that the stable region obtained by using nonlocal spin-transfer torque is smaller than
that by using its local counterpart.
This limitation is essential for multiple domain walls since the instability of
Gaussian wave packets will break the structure of multiple domain walls.
\end{abstract}


\maketitle

\section{Introduction.}
The dynamics of magnetic domain walls (DWs) in ferromagnetic
nanostrips has drawn considerable attention in the past few years.
An effective method to manipulate the DW is with electrical currents~\cite{Berger1986, Parkin2008,
Yan2010, Thiaville2005, Hayashi2007, Thomas2010, Franchin2011}.
The mechanism behind this is the so-called spin transfer torque (STT); when spin-polarized electrons
pass through the DW, electrons exert both adiabatic and nonadiabatic torques to the local magnetization~\cite{Zhang2004,Tatara2008}.
Various complex phenomena make DW motion interesting from a fundamental point of view.
For example, in thin nanostrips the Walker breakdown~\cite{Schryer1974} occurs when the
speed of a transverse DW reaches a critical velocity due to strong driving forces such as
external fields or spin-polarized currents. Hence, the Walker limit is the maximum velocity that a transverse DW can reach in thin strips (similar relativistic velocity limit can be found for antiferromagnetic DWs as well~\cite{Shiino2016}).
Interestingly, a transverse (head-to-head or tail-to-tail) DW does not suffer the Walker limit
in cylindrical nanowires~\cite{Yan2010, Goussev2010}; this is because the transverse DW in
cylindrical nanowires can rotate freely due to the absence of easy-plane anisotropy.
Therefore, we ask if there is a similar physical limit which determines the maximum velocity of the transverse DWs in cylindrical nanowires.

The Walker breakdown is related to the instability of domain walls, for example,
spin waves may be emitted when a transverse DW travels in a thin strip~\cite{Hu2013, Wang2014}.
Micromagnetic simulations show that a moving DW in cylindrical nanowires has an
almost vanishing mass~\cite{Yan2010, Hertel2015}.
It can be seen that the mass of the transverse DW is exactly zero for an arbitrary time-dependent charge
current when using the local spin transfer torque~\cite{Goussev2010}.
The local spin transfer torque $T_\mathrm{loc}$ has ignored the spin diffusion effect~\cite{Zhang2004, Claudio2012},
which is observable for sharp DWs. Moreover, when taking the spin flip into account,
the spin density does not simply depends on the local magnetization~\cite{Claudio2012}.
Therefore, it is interesting to ask whether the nonlocal spin transfer torque influences the stability of domain walls in cylindrical nanowires.

The dynamics of the magnetization in the presence of a spin polarized current is
governed by the extended Landau-Lifshitz-Gilbert (LLG) equation
with spin transfer torque $T$~\cite{Zhang2004, Thiaville2005, Tatara2008}:
\begin{equation}\label{eq_LLG}
\frac{\partial \mvec{m}}{\partial t}= -\gamma \mvec{m} \times \Heff
+\alpha\mvec{m} \times \frac{\partial \mvec{m}}{\partial t} - T
\end{equation}
where  $\mvec{m}$ is the unit vector of magnetization, $\gamma$ is the gyromagnetic ratio,
 $\Heff = -(1/\mu_0 M_s) \cdot {\delta E}/{\delta \mvec{m}}$
is the total effective field and $\alpha$ is the Gilbert damping. The local form of spin transfer torque reads
\begin{equation}\label{eq_T2}
T_\mathrm{loc} = u \partial_z \mvec{m} - \beta u \mvec{m} \times \partial_z \mvec{m}
\end{equation}
where the parameter $u=-j_z P g \mu_B /(2 e M_s)$ represents the strength of
spin-polarized current, and
$j_z$ is the charge current density along $z$ axis, $g$ is the Land\'{e} factor,
$\mu_B$ is the Bohr magneton, $P$ is the spin polarization rate, $e(>0)$ is the electron charge and $M_s$
is the saturation magnetization. The $\beta$ term is nonadiabatic torque, which influences
the spin waves amplitude significantly, leading to either a weakened or enhanced spin wave
attenuation depending on the relative direction between wave vector and charge current~\cite{Seo2009, Xia2016}.
The current-induced spin wave instability has been reported in Refs.~\cite{Bazaliy1998, Braun2004, Seo2009}.

The nonlocal (full version) spin transfer torque $T$ reads~\cite{Claudio2012}
\begin{equation}
T = \frac{1}{\tausd} \mvec{m} \times \delta \mvec{m}
\end{equation}
where $\delta \mvec{m}$ is the nonequilibrium spin density and $\tausd$ is $s$-$d$ exchange time.
The spin density $\delta \mvec{m}$ is obtained by solving the equation:
\begin{equation}\label{eq_T}
\frac{\partial \dm}{\partial t} = D \nabla^2 \dm + \frac{1}{\tausd} \mvec{m} \times \delta \mvec{m}
-\frac{1}{\tausf} \dm - u \partial_z \mvec{m}
\end{equation}
where $D$ is the diffusion constant, $\tausf$ is the spin-flip relaxation time.
Two characteristic lengths related to $\tausd$ and $\tausf$ can be defined:
$\lambda_\mathrm{J} = \sqrt{D \tausd}$ -- the diffusion length during the exchange time $\tausd$ and
$\lambda_\mathrm{sf} = \sqrt{D \tausf}$ --the diffusion length during the spin-flip time $\tausf$~\cite{Claudio2012}.
In the limit case $D=0$ and no time derivative, the torque $T$ reduces to the local torque $T_\mathrm{loc}$
and $\beta=\tausd/\tausf$.

\begin{figure}[tbhp]
\begin{center}
\includegraphics[width=0.48\textwidth]{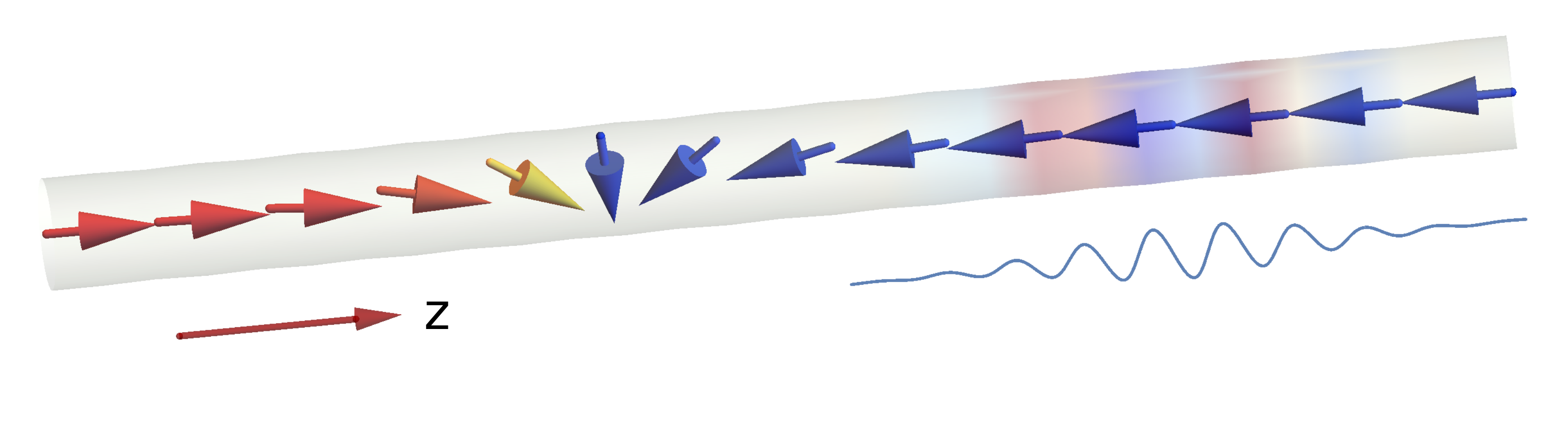}
\caption{Schematic illustration of a transverse (head-to-head) DW in a cylindrical nanowire. A Gaussian
wave packet will be generated when the charge current is applied suddenly.}
\label{fig_dw}
\end{center}
\end{figure}

The nonlocal spin transfer torque influences the vortex DW dynamics significantly~\cite{Claudio2012}.
However, it does not have a large influence for the transverse DWs.
In this paper, we will show that the nonlocal spin transfer torque induces spin wave (Gaussian wave packet)
emission in the presence of transverse domain wall when we apply the charge current to the system suddenly,
as sketched in Fig.~\ref{fig_dw}.

We consider a quasi-1D nanowire with the exchange interaction and an uniaxial anisotropy along the $z$ axis.
The total micromagnetic energy density of the system reads
\begin{equation}\label{eq_energy}
E = A (\nabla \mvec{m})^2 - K m_z^2
\end{equation}
where $A$ is the exchange constant, and $K$ is the anisotropy constant.
The demagnetization field is included in $K$ as an effective anisotropy for the cylindrical nanowire.

\section{Generation of Gaussian wave packets}
In this study, we perform micromagnetic simulation by solving the coupled LLG equation (\ref{eq_LLG})
and spin-density equation (\ref{eq_T}) simultaneously ~\cite{Wang}.
A fourth order accurate finite difference discretization in space is used to compute the effective fields
and the diffusion equation~(\ref{eq_T}).
We make use of the parameters of NiFe alloy~\cite{Claudio2012}: the exchange constant $A = 1 \times 10^{-11}\,\mathrm{J/m}$,
anisotropy $K=1\times 10^5\, \mathrm{J/m^3}$, the saturation magnetization
$M_s = 8\times 10^5\,\mathrm{A/m}$ and the damping coefficient  $\alpha=0.02$.
For given parameters, the domain wall width $\Delta=\sqrt{A/K}= 10$ nm. In the simulation,
the discretization size is chosen to be $\Delta z= 1$ nm.

\begin{figure}[tbhp]
\begin{center}
\includegraphics[width=0.45\textwidth]{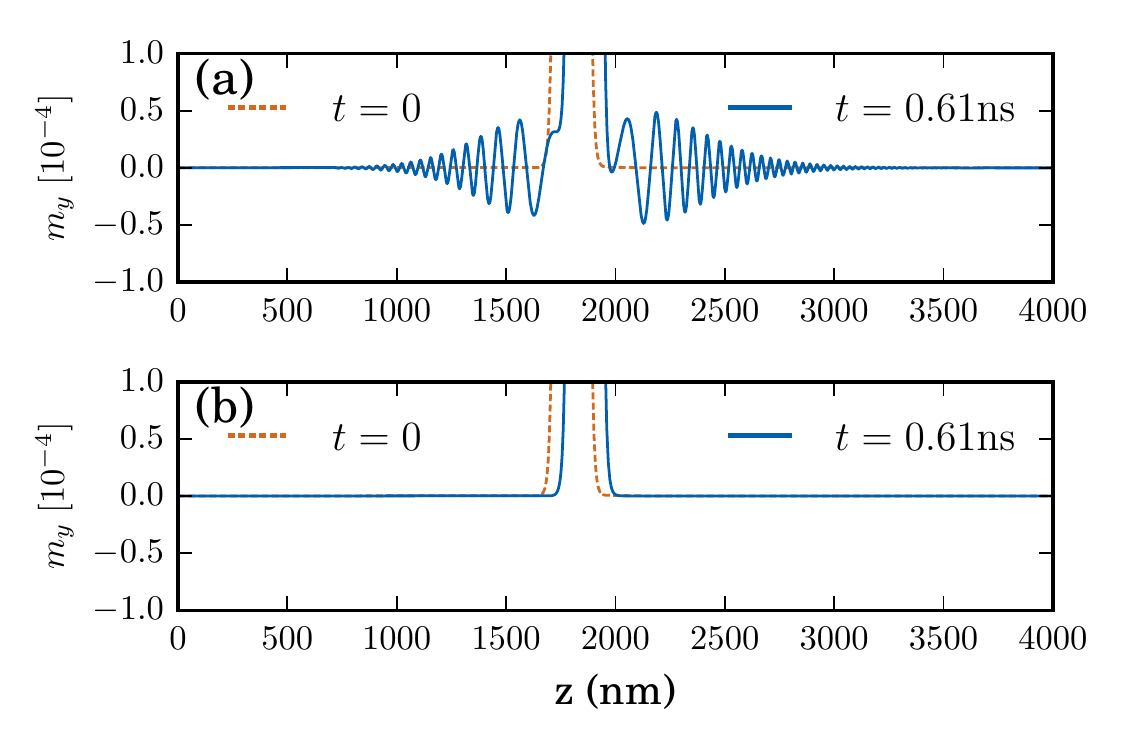}
\caption{(a) The snapshots of the $y$-component magnetization $m_y$
at $t=0$ ns and $t=0.61$ ns, where the nonlocal spin transfer torque is used.
(b) Same as (a) but the local spin transfer torque $T_\mathrm{local}$ is used.
}
\label{fig2}
\end{center}
\end{figure}

A head-to-head DW is placed in the middle of wire, as shown in Fig.~\ref{fig_dw}.
We apply a charge current in the $z$ direction with amplitude $u=100$ m/s,
and we can estimate that $u=100$ m/s corresponds
to a current density $j=1.972\times 10^{12}$ A/m$^2$ if $P=0.7$.
Interestingly, Gaussian wave packets are generated when we apply the charge current to the
system suddenly. Fig.~\ref{fig2}(a) shows the snapshots of the $y$-component magnetization $m_y$
at $t=0$ ns and $t=0.61$ ns, from which we can see the emergence of spin wave packets.
The parameters used in the simulation are $\lambda_\mathrm{sf}= 5$ nm, $\lambda_\mathrm{sf}= 1$ nm and $D=2.5\times 10^{-4}$ m$^2$/s.
As a comparison, we performed the micromagnetic simulation using the local spin transfer torque $T_\mathrm{loc}$.
Fig.~\ref{fig2}(b) shows the snapshots of $m_y$ at $t=0$ ns and $t=0.61$ ns, clearly, the DW moves smoothly without
spin waves emission.
The detailed animation [I.mp4] can be found in the Supplemental Material~\cite{Supp}.

The fact that no spin waves are emitted for the local spin transfer torque $T_\mathrm{loc}$ can be understood using
the Walker DW profile
\begin{equation}\label{eq_walker}
\theta_0(z)= 2 \arctan \exp (z/\Delta), \qquad \phi_0 = \mathrm{const}
\end{equation}
where we have written the magnetization unit vector $\mvec{m}$ as $\mvec{m}=(\sin \theta \cos \phi,
\sin \theta \sin \phi, \cos \theta)$. Eq.~(\ref{eq_walker}) describes a head-to-head domain wall
and $\Delta$ is the typical DW width.
The Walker solution for the LLG equation with $T_\mathrm{loc}$ is
\begin{equation}
\theta_*(z,t)=\theta_0(z-z_*(t)), \qquad \phi_*(z,t)=\phi_0(t),
\end{equation}
where~\cite{Wieser2010a, Yan2010}
\begin{equation}\label{eq_dw_v}
\dot{\phi}_*=\frac{(\beta-\alpha)u }{(1+\alpha^2)\Delta},
\qquad \dot{z}_*=\frac{(1+\alpha\beta)u }{1+\alpha^2}.
\end{equation}
Note that equation (\ref{eq_dw_v}) is an \textit{exact} spatiotemporal solution for the extended LLG equation
including the local STT, and thus the solution allows an arbitrary time-dependence function $u=u(t)$.
The procedure to show equation (\ref{eq_dw_v}) is an exact spatiotemporal solution can be found in the literature~\cite{Goussev2010}.

\begin{figure}[tbhp]
\begin{center}
\includegraphics[width=0.45\textwidth]{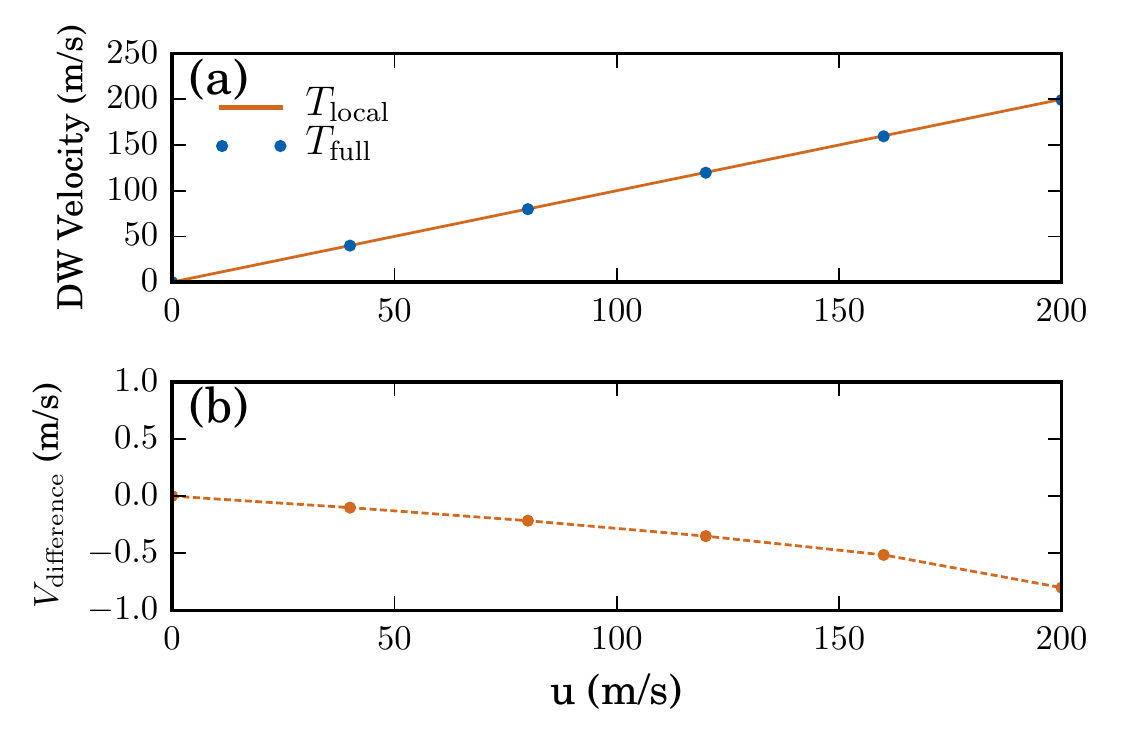}
\caption{
(a) The DW velocities as a function of $u$ for the both local ($T_\mathrm{local}$) and nonlocal ($T_\mathrm{full}$) cases.
(b) Plotting of DW velocity difference $V_\mathrm{difference}$ driven by local and nonlocal spin transfer torques, where
$\beta=0.04$ and $\alpha=0.02$ is used.
}
\label{fig3}
\end{center}
\end{figure}

The generation of spin waves only happens if we use the full version of spin transfer torques.
Moreover, the Walker solution is not an exact solution for the LLG equation with nonlocal spin transfer torques.
Fig.~\ref{fig3}(a) shows the stable DW velocities as a function of $u$ for the both local and nonlocal cases.
It is found that the DW velocity is very close to the Walker solution. Figure \ref{fig3}(b) plots the velocity
difference between the two cases, and the amplitude ratio is $\sim 0.5\%$.

\begin{figure}[tbhp]
\begin{center}
\includegraphics[width=0.45\textwidth]{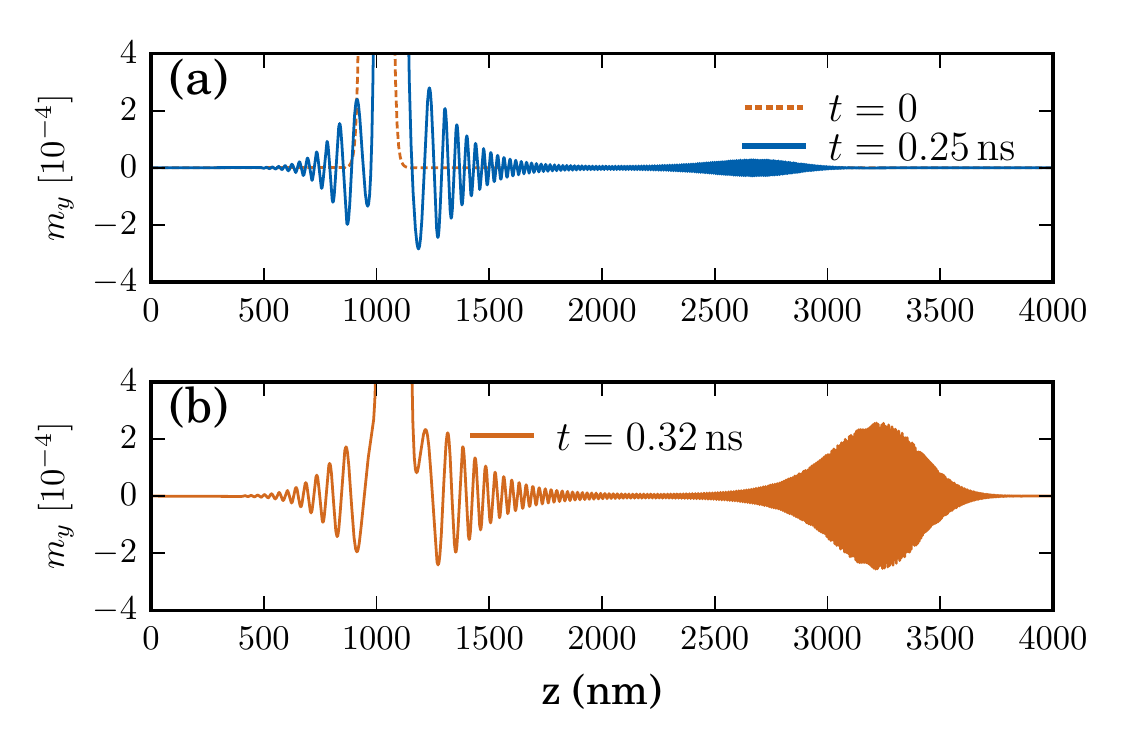}
\caption{Micromagnetic simulation results of the $y$-component magnetization $m_y$ at different time with $u=250$ m/s.
(a) The dashed line shows the initial DW profile at $t=0$. The blue line shows $m_y$ at $t=0.25$ ns.
(b) The snapshot of $m_y$ at $t=0.32$ ns -- the Gaussian packet is growing.}
\label{fig4}
\end{center}
\end{figure}

\section{Beyond the stable motion.}
From the animation [I.mp4], we can see that the amplitude of the wave packet is decreasing slowly,
where $u=100$ m/s is used.
There are two reasons responsible for the amplitude decrease: the existence of Gilbert damping and the intrinsic delocalization due to the quadratic dispersion relation of spin waves.

\begin{figure}[tbhp]
\begin{center}
\includegraphics[width=0.46\textwidth]{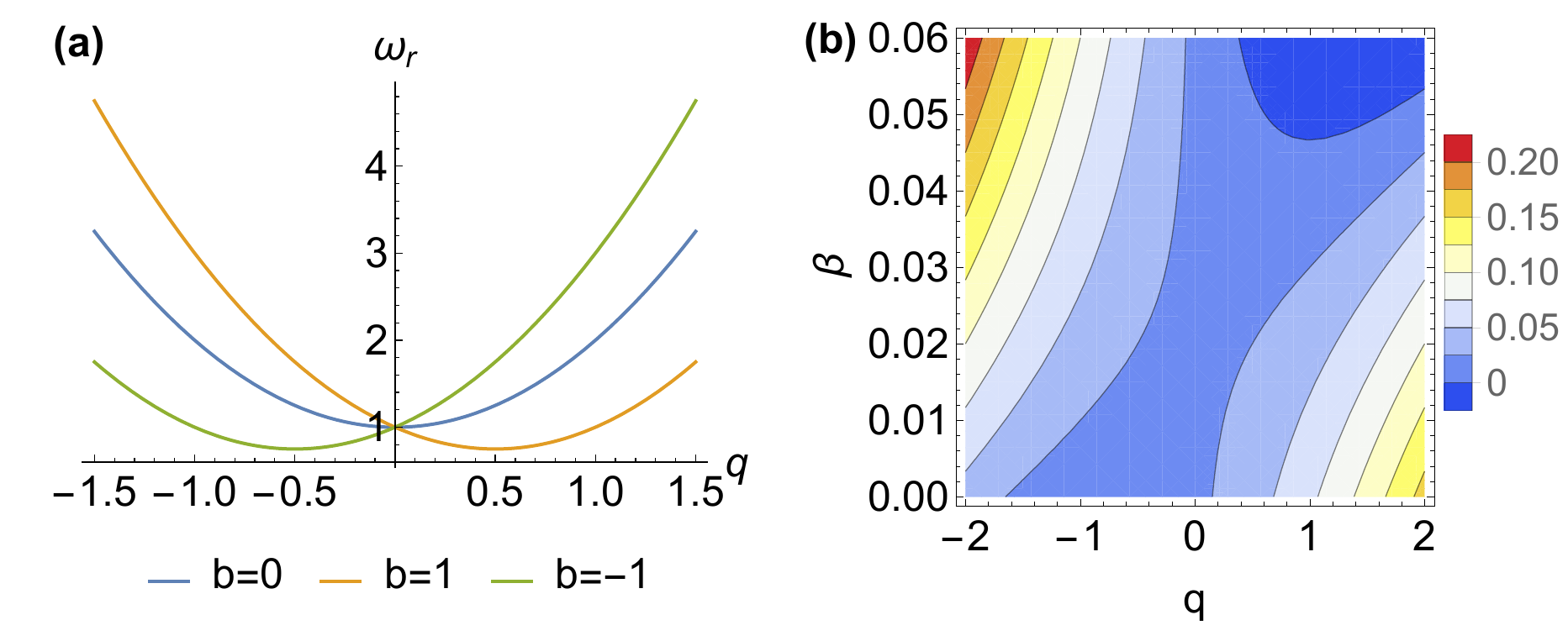}
\caption{(a) Dispersion relations for $b=0$, $b=-1$ and $b=1$. (b) Contour plot of the dissipation rate
$\omega_i$ as a function of $q$ and $\beta$ for $\alpha=0.02$ and $b=-1.5$.}
\label{fig_wr_wi}
\end{center}
\end{figure}

Fig.~\ref{fig4} shows the simulation results for the propagation of
the Gaussian wave packets using nonlocal spin transfer torque with $u=250$ m/s.
As can be seen by comparing Fig.~\ref{fig4}(a) and Fig.~\ref{fig4}(b), in this scenario
the Gaussian wave packet is growing. The corresponding animation [II.mp4] can be found in
Supplemental Material~\cite{Supp}. There are two panels in the animation show the same data
but with different scale.  The amplitude of the packet increases exponentially,
leading to a magnetization reversal and ending with a chaotic dynamics~\cite{Seo2009},

A chaotic dynamics of the domains gives an upper velocity limit for the DWs.
We now investigate the critical current density $u_c$ that determines whether the
Gaussian wave packet grows or decreases using the local spin transfer torque $T_\mathrm{local}$.
For the given energy density [Eq.~(\ref{eq_energy})], the corresponding effective field is
$\Heff= A_z (\partial^2 \mvec{m} /{\partial z^2}) +  K_z m_z \mvec{e}_{z}$ where $A_z = 2A/(\mu_0 M_s)$ and
$K_z = 2K/(\mu_0 M_s)$.
Substituting the effective field $\Heff$ into Eq.~(\ref{eq_LLG}), one obtains
\begin{eqnarray}\label{eq_LLG2}
\frac{\partial \mvec{m}}{\partial \tau} =  &- \mvec{m} \times \frac{\partial^2 \mvec{m}}{\partial \xi^2}
- m_z \mvec{m} \times \mvec{e}_z +\alpha \mvec{m} \times \frac{\partial \mvec{m}}{\partial \tau} \nonumber \\
&+ b \frac{ \partial  \mvec{m}}{\partial \xi} - \beta b \mvec{m} \times \frac{ \partial  \mvec{m}}{\partial \xi}
\end{eqnarray}
where $\tau = \gamma K_z t$, $\xi=z/\Delta$ and $b=-u/(\gamma K_z \Delta)$.
From Fig.~\ref{fig4}(b) we can find that the wave packet is far away from DW, so we
will consider the wave packet moves in uniform domains.
For example, the head-to-head DW separates two domains with magnetization $\mvec{m}=(0,0,1)$
and $\mvec{m}=(0,0,-1)$.
By introducing the complex transformation \cite{Li2007, Zhao2012}
\begin{equation}
\psi=m_x - i m_y \qquad m_z=\sqrt{1-|\psi|^2}
\end{equation}
and linearizing the LLG equation around $\mvec{m}=(0,0,1)$,
we arrive at a Schr\"{o}dinger-type equation
\begin{equation}\label{eq_schrodinger}
i (1+\alpha i)\frac{\partial \psi}{\partial \tau} = -  \frac{\partial^2 \psi }{\partial \xi^2} + \psi
+ i b (1+\beta i) \frac{\partial \psi }{\partial \xi}
\end{equation}
The complex conjugate of this corresponds to the linearized equation around $\mvec{m}=(0,0,-1)$
with transformation $m_z=-\sqrt{1-|\psi|^2}$.
A typical solution of Eq.~(\ref{eq_schrodinger}) is the travelling wave in the form $e^{i(q \xi- \omega\tau)}$
where $\xi$ and $\omega$ are dimensionless wave vector and frequency.
However, in the presence of damping (i.e., $\alpha>0$), a complex frequency $\tilde{\omega}$ or
wave vector $\tilde{q}$ must be introduced, which corresponds to a finite linewidth or
amplitude decay~\cite{Seo2009,Sekiguchi2012} of spin waves, respectively.
For spin waves with localized shape, we chose the former~\cite{Covington2002}, and thus
we look for a solution in the form
\begin{equation}\label{eq_gaussian}
\psi(\xi,\tau) = \frac{\psi_0}{\sqrt{2\pi}} \int_{-\infty}^{\infty} f(q) e^{i(q \xi- \tilde{\omega}\tau)} d q
\end{equation}
with $f(q)=e^{-a(q-q_0)^2}$ and $|\psi_0| \ll 1$. This solution represents a Gaussian wave packet.
Substituting Eq.~(\ref{eq_gaussian}) into Eq.~(\ref{eq_schrodinger}), we find
\begin{equation}
\tilde{\omega} = (\omega_r - i \omega_i)/(1+\alpha^2)
\end{equation}
where $\omega_r=(1+q^2-bq)-\alpha \beta b q$ gives the dispersion relation of spin waves
and $\omega_i=\alpha(1+q^2-bq)+\beta b q$ indicates the energy dissipation during the spin wave propagation.
Fig. \ref{fig_wr_wi}(a) shows the dispersion relations for different
values of $b$. We can see that the wave vector $q$ is shifted in the presence of spin current, and
the spin current is proportional to $b$~\cite{Bazaliy1998, Braun2004}. This wave vector shift is similar
to case that induced by the DMI~\cite{Moon2013}, resulting in an asymmetric spin-wave dispersion.
Note that the spin wave frequency $\omega_r$ is negative if $|b+\alpha \beta|>2$.
It is estimated that $b \sim 2$ corresponds to $u=879.3$ m/s for the
parameters we used above. The domain wall velocity $\dot{z}_*$ can be
obtained using Eq.~(\ref{eq_dw_v}) and is of similar magnitude.

Fig. \ref{fig_wr_wi}(b) shows a contour plot of the dissipation rate $\omega_i$ as a function
of $\beta$ and $q$ for $\alpha=0.02$ and $b=-0.15$, it is found that in the top right
corner $\omega_i$ is negative, which indicates that $\beta$ term may leads to a negative
dissipation rate if $|b|<2$.
For the case that $\alpha=0$ and $b=0$, the solution of $\psi(\xi,\tau)$ reduces to
\begin{equation}\label{eq_gaussian2}
\psi(\xi,\tau) = \sqrt{\frac{\psi_0^2/2}{a+i \tau}} e^{i[q_0 \xi-(1+q_0^2)\tau]
-(\xi-2 q_0\tau)^2/4(a+i \tau)}
\end{equation}
Eq.~(\ref{eq_gaussian2}) describes a moving Gaussian wave packet
with group velocity $v_g=(\partial \omega_r/\partial q)|_{q=q_0} = 2 q_0$. The packet delocalizes rapidly --
its amplitude decreases and width increases with time, similar to the one shown in Fig.~\ref{fig2}(b).
\begin{figure}[tbhp]
\begin{center}
\includegraphics[width=0.36\textwidth]{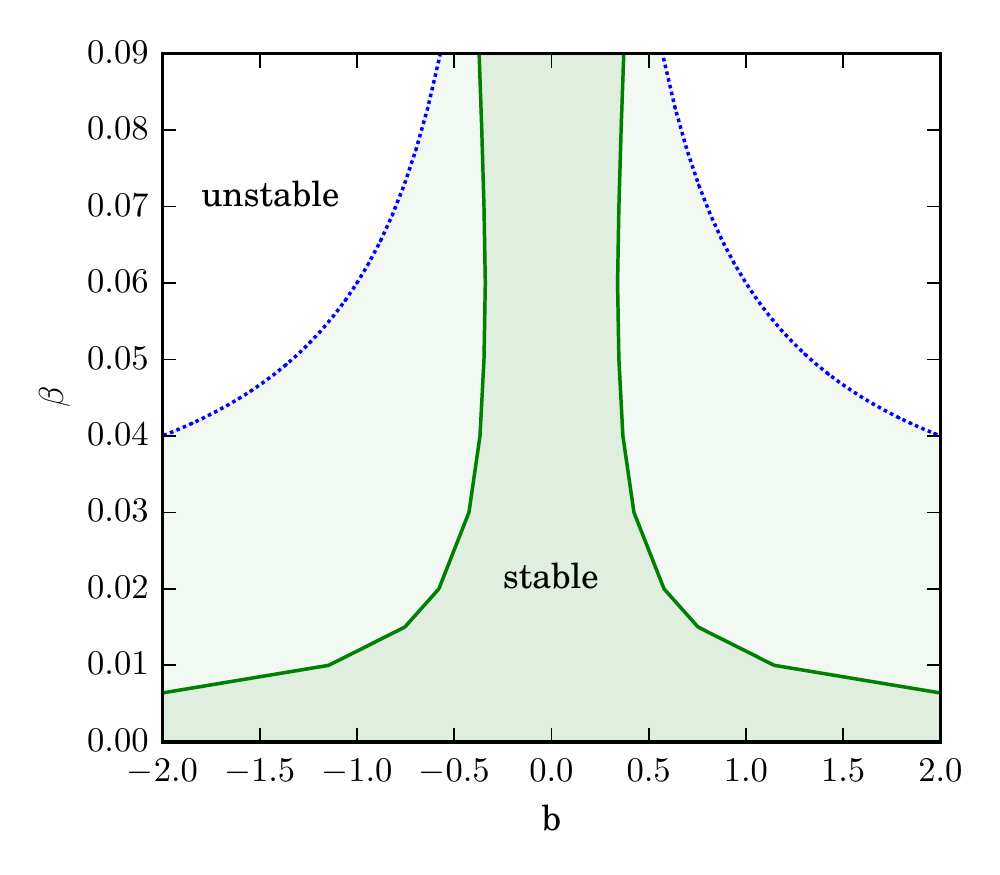}
\caption{The critical current density for $\alpha=0.02$. The dashed blue line is predicted using the
local spin transfer torque and the green line is obtained by using the micromagnetic simulation with
nonlocal spin transfer torque.}
\label{fig_phase}
\end{center}
\end{figure}

To analyze the stability of the wave packets, we define the total energy
$P(\tau) = (2\pi/\psi_0^2) \int_{-\infty}^\infty |\psi(\xi,\tau)|^2 d \xi$.
For the case $b>0$, it is convenient to compute $P$ in $q$ space, i.e., $P = \int_{-\infty}^\infty |\tilde{\psi}(q,\tau)|^2 dq$
where $\tilde{\psi}(q,\tau)= f(q)e^{-i \tilde{\omega} \tau}$
since Eq.~(\ref{eq_gaussian}) can be considered as a Fourier transformation of $\tilde{\psi}(q,\tau)$.
Hence, we obtain
\begin{equation}
P =
\sqrt{\frac{\pi/2}{a+\alpha \ttau}}
 \exp{\left[-\frac{g_0 \ttau^2+g_1 \ttau}{2(a+\alpha \ttau)}\right ]}
\end{equation}
where $\ttau=\tau/(1+\alpha^2)$, $g_0=4\alpha^2-(\alpha-\beta)^2b^2$ and
$g_1=4a[\alpha q_0^2 +b(\beta-\alpha) q_0+ \alpha]$.
It is clear that the sign of $g_0$ determines the total energy of the Gaussian wave packet over longer time scales;
a positive $g_0$ suppresses the Gaussian wave packet while a negative $g_0$ leads to a growing wave packet.
Therefore, the critical current density is determined by
\begin{equation}\label{eq_bc}
b_c =  \frac{2\alpha}{|\alpha-\beta|},
\end{equation}
which gives $u_c=2\gamma \sqrt{A_z K_z} \alpha /|\alpha-\beta|$ and thus the
critical current density $j_e^c$ reads
\begin{equation}\label{eq_jec}
j_e^c=   \frac{4e\sqrt{AK}}{ P \hbar} \frac{2\alpha}{|\alpha-\beta|}.
\end{equation}
It is perhaps surprising that this critical current density is independent from the group velocity $q_0$ and the Gaussian
wave width while for the spin wave case the amplitude decaying length is influenced by the wave vector of spin waves
for a given current density. Using the typical parameters of NiFe, one finds that $j_e^c$ is in the range of
 10$^{12}$--10$^{13}$ A/m$^2$. As a comparison, the critical $u$ for Walker breakdown in the presence of easy-plane anisotropy
$K_\perp$ is approximately established as $u_c^w= (1/4) u_c \kappa/\sqrt{1+\kappa/2}$ \cite{Thiaville2005}
where $\kappa=K_\perp/K_z$. For a typical $\kappa=0.5$ one obtains $u_c^w \approx 0.1 u_c$.

Fig.~\ref{fig_phase} plots the critical current density for $\alpha=0.02$.
The dashed blue line is plotted using Eq.~(\ref{eq_bc}), which is obtained
by checking the stability of the Gaussian wave packets using local spin transfer torque.
The green line is extracted from the micromagnetic simulation
with the nonlocal spin transfer torque. In detail, we monitor the maximum amplitude of the wave packet
for given charge current density $u$,  if the maximum amplitude always increase
we think the corresponding charge current $u$ is in the unstable region.
In the simulation, we fixed $\tausf=100$ fs and vary $\tausd$ according to the value of $\beta$.
As we can see, the stable region obtained using nonlocal spin transfer torque is much smaller than
that using its local counterpart. Moreover, for the local case,
when $\alpha=\beta$ the total energy of the Gaussian wave packet can be simplified to
$P = (\sqrt{\pi/2}/\sqrt{a+\alpha \ttau})
 e^{-2\alpha \ttau -2 a q_0^2 \alpha \ttau/(a+\alpha \ttau)}$.
We find that $P$ is independent from the charge current $b$ and decreases with $\ttau$ for arbitrary $q_0$,
which means that the wave packet is always suppressed if $\alpha=\beta$.
However, this conclusion is not true when we using the nonlocal spin transfer torque.

\section{Conclusion} In summary, we have studied the current-driven domain wall motion in a cylindrical nanowire
using micromagnetic simulation with nonlocal spin transfer torque.
We show that in the presence of domain wall a Gaussian wave packet will be generated
when the charge current is applied to the system suddenly.
The generation of wave packet only happens when we use the nonlocal spin transfer torque.
By analyzing the stability of the Gaussian wave packet, we give an upper velocity limit
for the domain wall motion driven with spin currents.
The limitation is especially important for multiple domain walls motion
since the instability of Gaussian wave packets will break their structure.
Moreover, the limitation also indicates that transverse domain walls can not move arbitrarily fast
in cylindrical nanowires though they are not subject to the Walker limit.

\section*{Acknowledgement}
We acknowledge the financial support from National Natural Science Foundation of China (Grants No.~11404280 and No.~11604169)
and EPSRC under Centre for Doctoral Training Grant EP/L015382/1. This work is sponsored by K.C.Wong Magna Fund in
Ningbo University.

\bibliographystyle{apsrev4-1}

\begin{thebibliography}{28}%
\makeatletter
\providecommand \@ifxundefined [1]{%
 \@ifx{#1\undefined}
}%
\providecommand \@ifnum [1]{%
 \ifnum #1\expandafter \@firstoftwo
 \else \expandafter \@secondoftwo
 \fi
}%
\providecommand \@ifx [1]{%
 \ifx #1\expandafter \@firstoftwo
 \else \expandafter \@secondoftwo
 \fi
}%
\providecommand \natexlab [1]{#1}%
\providecommand \enquote  [1]{``#1''}%
\providecommand \bibnamefont  [1]{#1}%
\providecommand \bibfnamefont [1]{#1}%
\providecommand \citenamefont [1]{#1}%
\providecommand \href@noop [0]{\@secondoftwo}%
\providecommand \href [0]{\begingroup \@sanitize@url \@href}%
\providecommand \@href[1]{\@@startlink{#1}\@@href}%
\providecommand \@@href[1]{\endgroup#1\@@endlink}%
\providecommand \@sanitize@url [0]{\catcode `\\12\catcode `\$12\catcode
  `\&12\catcode `\#12\catcode `\^12\catcode `\_12\catcode `\%12\relax}%
\providecommand \@@startlink[1]{}%
\providecommand \@@endlink[0]{}%
\providecommand \url  [0]{\begingroup\@sanitize@url \@url }%
\providecommand \@url [1]{\endgroup\@href {#1}{\urlprefix }}%
\providecommand \urlprefix  [0]{URL }%
\providecommand \Eprint [0]{\href }%
\providecommand \doibase [0]{http://dx.doi.org/}%
\providecommand \selectlanguage [0]{\@gobble}%
\providecommand \bibinfo  [0]{\@secondoftwo}%
\providecommand \bibfield  [0]{\@secondoftwo}%
\providecommand \translation [1]{[#1]}%
\providecommand \BibitemOpen [0]{}%
\providecommand \bibitemStop [0]{}%
\providecommand \bibitemNoStop [0]{.\EOS\space}%
\providecommand \EOS [0]{\spacefactor3000\relax}%
\providecommand \BibitemShut  [1]{\csname bibitem#1\endcsname}%
\let\auto@bib@innerbib\@empty
\bibitem [{\citenamefont {Berger}(1986)}]{Berger1986}%
  \BibitemOpen
  \bibfield  {author} {\bibinfo {author} {\bibfnamefont {L.}~\bibnamefont
  {Berger}},\ }\href {\doibase 10.1103/PhysRevB.33.1572} {\bibfield  {journal}
  {\bibinfo  {journal} {Phys. Rev. B}\ }\textbf {\bibinfo {volume} {33}},\
  \bibinfo {pages} {1572} (\bibinfo {year} {1986})}\BibitemShut {NoStop}%
\bibitem [{\citenamefont {Parkin}\ \emph {et~al.}(2008)\citenamefont {Parkin},
  \citenamefont {Hayashi},\ and\ \citenamefont {Thomas}}]{Parkin2008}%
  \BibitemOpen
  \bibfield  {author} {\bibinfo {author} {\bibfnamefont {S.~S.~P.}\
  \bibnamefont {Parkin}}, \bibinfo {author} {\bibfnamefont {M.}~\bibnamefont
  {Hayashi}}, \ and\ \bibinfo {author} {\bibfnamefont {L.}~\bibnamefont
  {Thomas}},\ }\href {\doibase 10.1126/science.1145799} {\bibfield  {journal}
  {\bibinfo  {journal} {Science}\ }\textbf {\bibinfo {volume} {320}},\ \bibinfo
  {pages} {190} (\bibinfo {year} {2008})}\BibitemShut {NoStop}%
\bibitem [{\citenamefont {Yan}\ \emph {et~al.}(2010)\citenamefont {Yan},
  \citenamefont {K{\'{a}}kay}, \citenamefont {Gliga},\ and\ \citenamefont
  {Hertel}}]{Yan2010}%
  \BibitemOpen
  \bibfield  {author} {\bibinfo {author} {\bibfnamefont {M.}~\bibnamefont
  {Yan}}, \bibinfo {author} {\bibfnamefont {A.}~\bibnamefont {K{\'{a}}kay}},
  \bibinfo {author} {\bibfnamefont {S.}~\bibnamefont {Gliga}}, \ and\ \bibinfo
  {author} {\bibfnamefont {R.}~\bibnamefont {Hertel}},\ }\href {\doibase
  10.1103/PhysRevLett.104.057201} {\bibfield  {journal} {\bibinfo  {journal}
  {Phys. Rev. Lett.}\ }\textbf {\bibinfo {volume} {104}},\ \bibinfo {pages}
  {057201} (\bibinfo {year} {2010})}\BibitemShut {NoStop}%
\bibitem [{\citenamefont {Thiaville}\ \emph {et~al.}(2005)\citenamefont
  {Thiaville}, \citenamefont {Nakatani}, \citenamefont {Miltat},\ and\
  \citenamefont {Suzuki}}]{Thiaville2005}%
  \BibitemOpen
  \bibfield  {author} {\bibinfo {author} {\bibfnamefont {A.}~\bibnamefont
  {Thiaville}}, \bibinfo {author} {\bibfnamefont {Y.}~\bibnamefont {Nakatani}},
  \bibinfo {author} {\bibfnamefont {J.}~\bibnamefont {Miltat}}, \ and\ \bibinfo
  {author} {\bibfnamefont {Y.}~\bibnamefont {Suzuki}},\ }\href {\doibase
  10.1209/epl/i2004-10452-6} {\bibfield  {journal} {\bibinfo  {journal}
  {Europhys. Lett.}\ }\textbf {\bibinfo {volume} {69}},\ \bibinfo {pages} {990}
  (\bibinfo {year} {2005})}\BibitemShut {NoStop}%
\bibitem [{\citenamefont {Hayashi}\ \emph {et~al.}(2007)\citenamefont
  {Hayashi}, \citenamefont {Thomas}, \citenamefont {Rettner}, \citenamefont
  {Moriya}, \citenamefont {Bazaliy},\ and\ \citenamefont
  {Parkin}}]{Hayashi2007}%
  \BibitemOpen
  \bibfield  {author} {\bibinfo {author} {\bibfnamefont {M.}~\bibnamefont
  {Hayashi}}, \bibinfo {author} {\bibfnamefont {L.}~\bibnamefont {Thomas}},
  \bibinfo {author} {\bibfnamefont {C.}~\bibnamefont {Rettner}}, \bibinfo
  {author} {\bibfnamefont {R.}~\bibnamefont {Moriya}}, \bibinfo {author}
  {\bibfnamefont {Y.~B.}\ \bibnamefont {Bazaliy}}, \ and\ \bibinfo {author}
  {\bibfnamefont {S.~S.~P.}\ \bibnamefont {Parkin}},\ }\href {\doibase
  10.1103/PhysRevLett.98.037204} {\bibfield  {journal} {\bibinfo  {journal}
  {Phys. Rev. Lett.}\ }\textbf {\bibinfo {volume} {98}},\ \bibinfo {pages}
  {037204} (\bibinfo {year} {2007})}\BibitemShut {NoStop}%
\bibitem [{\citenamefont {Thomas}\ \emph {et~al.}(2010)\citenamefont {Thomas},
  \citenamefont {Moriya}, \citenamefont {Rettner},\ and\ \citenamefont
  {Parkin}}]{Thomas2010}%
  \BibitemOpen
  \bibfield  {author} {\bibinfo {author} {\bibfnamefont {L.}~\bibnamefont
  {Thomas}}, \bibinfo {author} {\bibfnamefont {R.}~\bibnamefont {Moriya}},
  \bibinfo {author} {\bibfnamefont {C.}~\bibnamefont {Rettner}}, \ and\
  \bibinfo {author} {\bibfnamefont {S.~S.~P.}\ \bibnamefont {Parkin}},\ }\href
  {\doibase 10.1126/science.1197468} {\bibfield  {journal} {\bibinfo  {journal}
  {Science}\ }\textbf {\bibinfo {volume} {330}},\ \bibinfo {pages} {1810}
  (\bibinfo {year} {2010})}\BibitemShut {NoStop}%
\bibitem [{\citenamefont {Franchin}\ \emph {et~al.}(2011)\citenamefont
  {Franchin}, \citenamefont {Knittel}, \citenamefont {Albert}, \citenamefont
  {Chernyshenko}, \citenamefont {Fischbacher}, \citenamefont {Prabhakar},\ and\
  \citenamefont {Fangohr}}]{Franchin2011}%
  \BibitemOpen
  \bibfield  {author} {\bibinfo {author} {\bibfnamefont {M.}~\bibnamefont
  {Franchin}}, \bibinfo {author} {\bibfnamefont {A.}~\bibnamefont {Knittel}},
  \bibinfo {author} {\bibfnamefont {M.}~\bibnamefont {Albert}}, \bibinfo
  {author} {\bibfnamefont {D.~S.}\ \bibnamefont {Chernyshenko}}, \bibinfo
  {author} {\bibfnamefont {T.}~\bibnamefont {Fischbacher}}, \bibinfo {author}
  {\bibfnamefont {A.}~\bibnamefont {Prabhakar}}, \ and\ \bibinfo {author}
  {\bibfnamefont {H.}~\bibnamefont {Fangohr}},\ }\href {\doibase
  10.1103/PhysRevB.84.094409} {\bibfield  {journal} {\bibinfo  {journal} {Phys.
  Rev. B}\ }\textbf {\bibinfo {volume} {84}},\ \bibinfo {pages} {094409}
  (\bibinfo {year} {2011})},\ \Eprint {http://arxiv.org/abs/1104.3010}
  {1104.3010} \BibitemShut {NoStop}%
\bibitem [{\citenamefont {Zhang}\ and\ \citenamefont {Li}(2004)}]{Zhang2004}%
  \BibitemOpen
  \bibfield  {author} {\bibinfo {author} {\bibfnamefont {S.}~\bibnamefont
  {Zhang}}\ and\ \bibinfo {author} {\bibfnamefont {Z.}~\bibnamefont {Li}},\
  }\href {\doibase 10.1103/PhysRevLett.93.127204} {\bibfield  {journal}
  {\bibinfo  {journal} {Phys. Rev. Lett.}\ }\textbf {\bibinfo {volume} {93}},\
  \bibinfo {pages} {127204} (\bibinfo {year} {2004})}\BibitemShut {NoStop}%
\bibitem [{\citenamefont {Tatara}\ \emph {et~al.}(2008)\citenamefont {Tatara},
  \citenamefont {Kohno},\ and\ \citenamefont {Shibata}}]{Tatara2008}%
  \BibitemOpen
  \bibfield  {author} {\bibinfo {author} {\bibfnamefont {G.}~\bibnamefont
  {Tatara}}, \bibinfo {author} {\bibfnamefont {H.}~\bibnamefont {Kohno}}, \
  and\ \bibinfo {author} {\bibfnamefont {J.}~\bibnamefont {Shibata}},\ }\href
  {\doibase 10.1016/j.physrep.2008.07.003} {\bibfield  {journal} {\bibinfo
  {journal} {Phys. Rep.}\ }\textbf {\bibinfo {volume} {468}},\ \bibinfo {pages}
  {213} (\bibinfo {year} {2008})}\BibitemShut {NoStop}%
\bibitem [{\citenamefont {Schryer}\ and\ \citenamefont
  {Walker}(1974)}]{Schryer1974}%
  \BibitemOpen
  \bibfield  {author} {\bibinfo {author} {\bibfnamefont {N.~L.}\ \bibnamefont
  {Schryer}}\ and\ \bibinfo {author} {\bibfnamefont {L.~R.}\ \bibnamefont
  {Walker}},\ }\href {\doibase 10.1063/1.1663252} {\bibfield  {journal}
  {\bibinfo  {journal} {J. Appl. Phys.}\ }\textbf {\bibinfo {volume} {45}},\
  \bibinfo {pages} {5406} (\bibinfo {year} {1974})}\BibitemShut {NoStop}%
\bibitem [{\citenamefont {Shiino}\ \emph {et~al.}(2016)\citenamefont {Shiino},
  \citenamefont {Oh}, \citenamefont {Haney}, \citenamefont {Lee}, \citenamefont
  {Go}, \citenamefont {Park},\ and\ \citenamefont {Lee}}]{Shiino2016}%
  \BibitemOpen
  \bibfield  {author} {\bibinfo {author} {\bibfnamefont {T.}~\bibnamefont
  {Shiino}}, \bibinfo {author} {\bibfnamefont {S.-H.}\ \bibnamefont {Oh}},
  \bibinfo {author} {\bibfnamefont {P.~M.}\ \bibnamefont {Haney}}, \bibinfo
  {author} {\bibfnamefont {S.-W.}\ \bibnamefont {Lee}}, \bibinfo {author}
  {\bibfnamefont {G.}~\bibnamefont {Go}}, \bibinfo {author} {\bibfnamefont
  {B.-G.}\ \bibnamefont {Park}}, \ and\ \bibinfo {author} {\bibfnamefont
  {K.-J.}\ \bibnamefont {Lee}},\ }\href {\doibase
  10.1103/PhysRevLett.117.087203} {\bibfield  {journal} {\bibinfo  {journal}
  {Phys. Rev. Lett.}\ }\textbf {\bibinfo {volume} {117}},\ \bibinfo {pages}
  {087203} (\bibinfo {year} {2016})},\ \Eprint
  {http://arxiv.org/abs/1604.01473} {1604.01473} \BibitemShut {NoStop}%
\bibitem [{\citenamefont {Goussev}\ \emph {et~al.}(2010)\citenamefont
  {Goussev}, \citenamefont {Robbins},\ and\ \citenamefont
  {Slastikov}}]{Goussev2010}%
  \BibitemOpen
  \bibfield  {author} {\bibinfo {author} {\bibfnamefont {A.}~\bibnamefont
  {Goussev}}, \bibinfo {author} {\bibfnamefont {J.}~\bibnamefont {Robbins}}, \
  and\ \bibinfo {author} {\bibfnamefont {V.}~\bibnamefont {Slastikov}},\ }\href
  {\doibase 10.1103/PhysRevLett.104.147202} {\bibfield  {journal} {\bibinfo
  {journal} {Phys. Rev. Lett.}\ }\textbf {\bibinfo {volume} {104}},\ \bibinfo
  {pages} {147202} (\bibinfo {year} {2010})}\BibitemShut {NoStop}%
\bibitem [{\citenamefont {Hu}\ and\ \citenamefont {Wang}(2013)}]{Hu2013}%
  \BibitemOpen
  \bibfield  {author} {\bibinfo {author} {\bibfnamefont {B.}~\bibnamefont
  {Hu}}\ and\ \bibinfo {author} {\bibfnamefont {X.~R.}\ \bibnamefont {Wang}},\
  }\href {\doibase 10.1103/PhysRevLett.111.027205} {\bibfield  {journal}
  {\bibinfo  {journal} {Phys. Rev. Lett.}\ }\textbf {\bibinfo {volume} {111}},\
  \bibinfo {pages} {027205} (\bibinfo {year} {2013})}\BibitemShut {NoStop}%
\bibitem [{\citenamefont {Wang}\ and\ \citenamefont {Wang}(2014)}]{Wang2014}%
  \BibitemOpen
  \bibfield  {author} {\bibinfo {author} {\bibfnamefont {X.~S.}\ \bibnamefont
  {Wang}}\ and\ \bibinfo {author} {\bibfnamefont {X.~R.}\ \bibnamefont
  {Wang}},\ }\href {\doibase 10.1103/PhysRevB.90.184415} {\bibfield  {journal}
  {\bibinfo  {journal} {Phys. Rev. B}\ }\textbf {\bibinfo {volume} {90}},\
  \bibinfo {pages} {184415} (\bibinfo {year} {2014})}\BibitemShut {NoStop}%
\bibitem [{\citenamefont {Hertel}\ and\ \citenamefont
  {K{\'{a}}kay}(2015)}]{Hertel2015}%
  \BibitemOpen
  \bibfield  {author} {\bibinfo {author} {\bibfnamefont {R.}~\bibnamefont
  {Hertel}}\ and\ \bibinfo {author} {\bibfnamefont {A.}~\bibnamefont
  {K{\'{a}}kay}},\ }\href {\doibase 10.1016/j.jmmm.2014.11.073} {\bibfield
  {journal} {\bibinfo  {journal} {J. Magn. Magn. Mater.}\ }\textbf {\bibinfo
  {volume} {379}},\ \bibinfo {pages} {45} (\bibinfo {year} {2015})}\BibitemShut
  {NoStop}%
\bibitem [{\citenamefont {Claudio-Gonzalez}\ \emph {et~al.}(2012)\citenamefont
  {Claudio-Gonzalez}, \citenamefont {Thiaville},\ and\ \citenamefont
  {Miltat}}]{Claudio2012}%
  \BibitemOpen
  \bibfield  {author} {\bibinfo {author} {\bibfnamefont {D.}~\bibnamefont
  {Claudio-Gonzalez}}, \bibinfo {author} {\bibfnamefont {A.}~\bibnamefont
  {Thiaville}}, \ and\ \bibinfo {author} {\bibfnamefont {J.}~\bibnamefont
  {Miltat}},\ }\href {\doibase 10.1103/PhysRevLett.108.227208} {\bibfield
  {journal} {\bibinfo  {journal} {Phys. Rev. Lett.}\ }\textbf {\bibinfo
  {volume} {108}},\ \bibinfo {pages} {227208} (\bibinfo {year}
  {2012})}\BibitemShut {NoStop}%
\bibitem [{\citenamefont {Seo}\ \emph {et~al.}(2009)\citenamefont {Seo},
  \citenamefont {Lee}, \citenamefont {Yang},\ and\ \citenamefont
  {Ono}}]{Seo2009}%
  \BibitemOpen
  \bibfield  {author} {\bibinfo {author} {\bibfnamefont {S.-M.}\ \bibnamefont
  {Seo}}, \bibinfo {author} {\bibfnamefont {K.-J.}\ \bibnamefont {Lee}},
  \bibinfo {author} {\bibfnamefont {H.}~\bibnamefont {Yang}}, \ and\ \bibinfo
  {author} {\bibfnamefont {T.}~\bibnamefont {Ono}},\ }\href {\doibase
  10.1103/PhysRevLett.102.147202} {\bibfield  {journal} {\bibinfo  {journal}
  {Phys. Rev. Lett.}\ }\textbf {\bibinfo {volume} {102}},\ \bibinfo {pages}
  {147202} (\bibinfo {year} {2009})}\BibitemShut {NoStop}%
\bibitem [{\citenamefont {Xia}\ \emph {et~al.}(2016)\citenamefont {Xia},
  \citenamefont {Chen}, \citenamefont {Zeng},\ and\ \citenamefont
  {Yan}}]{Xia2016}%
  \BibitemOpen
  \bibfield  {author} {\bibinfo {author} {\bibfnamefont {H.}~\bibnamefont
  {Xia}}, \bibinfo {author} {\bibfnamefont {J.}~\bibnamefont {Chen}}, \bibinfo
  {author} {\bibfnamefont {X.}~\bibnamefont {Zeng}}, \ and\ \bibinfo {author}
  {\bibfnamefont {M.}~\bibnamefont {Yan}},\ }\href {\doibase
  10.1103/PhysRevB.93.140410} {\bibfield  {journal} {\bibinfo  {journal} {Phys.
  Rev. B}\ }\textbf {\bibinfo {volume} {93}},\ \bibinfo {pages} {140410}
  (\bibinfo {year} {2016})}\BibitemShut {NoStop}%
\bibitem [{\citenamefont {Bazaliy}\ \emph {et~al.}(1998)\citenamefont
  {Bazaliy}, \citenamefont {Jones},\ and\ \citenamefont {Zhang}}]{Bazaliy1998}%
  \BibitemOpen
  \bibfield  {author} {\bibinfo {author} {\bibfnamefont {Y.}~\bibnamefont
  {Bazaliy}}, \bibinfo {author} {\bibfnamefont {B.}~\bibnamefont {Jones}}, \
  and\ \bibinfo {author} {\bibfnamefont {S.-C.}\ \bibnamefont {Zhang}},\ }\href
  {\doibase 10.1103/PhysRevB.57.R3213} {\bibfield  {journal} {\bibinfo
  {journal} {Phys. Rev. B}\ }\textbf {\bibinfo {volume} {57}},\ \bibinfo
  {pages} {R3213} (\bibinfo {year} {1998})}\BibitemShut {NoStop}%
\bibitem [{\citenamefont {Fern\'andez-Rossier}\ \emph
  {et~al.}(2004)\citenamefont {Fern\'andez-Rossier}, \citenamefont {Braun},
  \citenamefont {N\'u\~nez},\ and\ \citenamefont {MacDonald}}]{Braun2004}%
  \BibitemOpen
  \bibfield  {author} {\bibinfo {author} {\bibfnamefont {J.}~\bibnamefont
  {Fern\'andez-Rossier}}, \bibinfo {author} {\bibfnamefont {M.}~\bibnamefont
  {Braun}}, \bibinfo {author} {\bibfnamefont {A.~S.}\ \bibnamefont
  {N\'u\~nez}}, \ and\ \bibinfo {author} {\bibfnamefont {A.~H.}\ \bibnamefont
  {MacDonald}},\ }\href {\doibase 10.1103/PhysRevB.69.174412} {\bibfield
  {journal} {\bibinfo  {journal} {Phys. Rev. B}\ }\textbf {\bibinfo {volume}
  {69}},\ \bibinfo {pages} {174412} (\bibinfo {year} {2004})}\BibitemShut
  {NoStop}%
\bibitem [{\citenamefont {Wang}\ \emph {et~al.}()\citenamefont {Wang},
  \citenamefont {Cortes}, \citenamefont {Vousden}, \citenamefont {Carey},
  \citenamefont {Beg},\ and\ \citenamefont {Fangohr}}]{Wang}%
  \BibitemOpen
  \bibfield  {author} {\bibinfo {author} {\bibfnamefont {W.}~\bibnamefont
  {Wang}}, \bibinfo {author} {\bibfnamefont {M.-A. B.~D.}\ \bibnamefont
  {Cortes}}, \bibinfo {author} {\bibfnamefont {M.}~\bibnamefont {Vousden}},
  \bibinfo {author} {\bibfnamefont {B.}~\bibnamefont {Carey}}, \bibinfo
  {author} {\bibfnamefont {M.}~\bibnamefont {Beg}}, \ and\ \bibinfo {author}
  {\bibfnamefont {H.}~\bibnamefont {Fangohr}},\ }\href@noop {} {}\bibinfo
  {note} {Fidimag,
  \url{http://computationalmodelling.github.io/fidimag/}}\BibitemShut {NoStop}%
\bibitem [{Sup()}]{Supp}%
  \BibitemOpen
  \href@noop {} {}\bibinfo {note} {See supplementary material at [URL] for the
  animations of the creation and propagation of Gaussian wave
  packets.}\BibitemShut {Stop}%
\bibitem [{\citenamefont {Wieser}\ \emph {et~al.}(2010)\citenamefont {Wieser},
  \citenamefont {Vedmedenko},\ and\ \citenamefont
  {Wiesendanger}}]{Wieser2010a}%
  \BibitemOpen
  \bibfield  {author} {\bibinfo {author} {\bibfnamefont {R.}~\bibnamefont
  {Wieser}}, \bibinfo {author} {\bibfnamefont {E.~Y.}\ \bibnamefont
  {Vedmedenko}}, \ and\ \bibinfo {author} {\bibfnamefont {R.}~\bibnamefont
  {Wiesendanger}},\ }\href {\doibase 10.1103/PhysRevB.81.024405} {\bibfield
  {journal} {\bibinfo  {journal} {Phys. Rev. B}\ }\textbf {\bibinfo {volume}
  {81}},\ \bibinfo {pages} {024405} (\bibinfo {year} {2010})}\BibitemShut
  {NoStop}%
\bibitem [{\citenamefont {Li}\ \emph {et~al.}(2007)\citenamefont {Li},
  \citenamefont {Li}, \citenamefont {Li},\ and\ \citenamefont {Liu}}]{Li2007}%
  \BibitemOpen
  \bibfield  {author} {\bibinfo {author} {\bibfnamefont {Z.~D.}\ \bibnamefont
  {Li}}, \bibinfo {author} {\bibfnamefont {Q.~Y.}\ \bibnamefont {Li}}, \bibinfo
  {author} {\bibfnamefont {L.}~\bibnamefont {Li}}, \ and\ \bibinfo {author}
  {\bibfnamefont {W.~M.}\ \bibnamefont {Liu}},\ }\href {\doibase
  10.1103/PhysRevE.76.026605} {\bibfield  {journal} {\bibinfo  {journal} {Phys.
  Rev. E}\ }\textbf {\bibinfo {volume} {76}},\ \bibinfo {pages} {026605}
  (\bibinfo {year} {2007})}\BibitemShut {NoStop}%
\bibitem [{\citenamefont {Zhao}\ \emph {et~al.}(2012)\citenamefont {Zhao},
  \citenamefont {Li}, \citenamefont {Li}, \citenamefont {Wen}, \citenamefont
  {Fu},\ and\ \citenamefont {Liu}}]{Zhao2012}%
  \BibitemOpen
  \bibfield  {author} {\bibinfo {author} {\bibfnamefont {F.}~\bibnamefont
  {Zhao}}, \bibinfo {author} {\bibfnamefont {Z.~D.}\ \bibnamefont {Li}},
  \bibinfo {author} {\bibfnamefont {Q.~Y.}\ \bibnamefont {Li}}, \bibinfo
  {author} {\bibfnamefont {L.}~\bibnamefont {Wen}}, \bibinfo {author}
  {\bibfnamefont {G.}~\bibnamefont {Fu}}, \ and\ \bibinfo {author}
  {\bibfnamefont {W.~M.}\ \bibnamefont {Liu}},\ }\href {\doibase
  10.1016/j.aop.2012.05.012} {\bibfield  {journal} {\bibinfo  {journal} {Ann.
  Phys. (N.Y.)}\ }\textbf {\bibinfo {volume} {327}},\ \bibinfo {pages} {2085}
  (\bibinfo {year} {2012})}\BibitemShut {NoStop}%
\bibitem [{\citenamefont {Sekiguchi}\ \emph {et~al.}(2012)\citenamefont
  {Sekiguchi}, \citenamefont {Yamada}, \citenamefont {Seo}, \citenamefont
  {Lee}, \citenamefont {Chiba}, \citenamefont {Kobayashi},\ and\ \citenamefont
  {Ono}}]{Sekiguchi2012}%
  \BibitemOpen
  \bibfield  {author} {\bibinfo {author} {\bibfnamefont {K.}~\bibnamefont
  {Sekiguchi}}, \bibinfo {author} {\bibfnamefont {K.}~\bibnamefont {Yamada}},
  \bibinfo {author} {\bibfnamefont {S.~M.}\ \bibnamefont {Seo}}, \bibinfo
  {author} {\bibfnamefont {K.~J.}\ \bibnamefont {Lee}}, \bibinfo {author}
  {\bibfnamefont {D.}~\bibnamefont {Chiba}}, \bibinfo {author} {\bibfnamefont
  {K.}~\bibnamefont {Kobayashi}}, \ and\ \bibinfo {author} {\bibfnamefont
  {T.}~\bibnamefont {Ono}},\ }\href {\doibase 10.1103/PhysRevLett.108.017203}
  {\bibfield  {journal} {\bibinfo  {journal} {Phys. Rev. Lett.}\ }\textbf
  {\bibinfo {volume} {108}},\ \bibinfo {pages} {017203} (\bibinfo {year}
  {2012})}\BibitemShut {NoStop}%
\bibitem [{\citenamefont {Covington}\ \emph {et~al.}(2002)\citenamefont
  {Covington}, \citenamefont {Crawford},\ and\ \citenamefont
  {Parker}}]{Covington2002}%
  \BibitemOpen
  \bibfield  {author} {\bibinfo {author} {\bibfnamefont {M.}~\bibnamefont
  {Covington}}, \bibinfo {author} {\bibfnamefont {T.~M.}\ \bibnamefont
  {Crawford}}, \ and\ \bibinfo {author} {\bibfnamefont {G.~J.}\ \bibnamefont
  {Parker}},\ }\href {\doibase 10.1103/PhysRevLett.89.237202} {\bibfield
  {journal} {\bibinfo  {journal} {Phys. Rev. Lett.}\ }\textbf {\bibinfo
  {volume} {89}},\ \bibinfo {pages} {237202} (\bibinfo {year}
  {2002})}\BibitemShut {NoStop}%
\bibitem [{\citenamefont {Moon}\ \emph {et~al.}(2013)\citenamefont {Moon},
  \citenamefont {Seo}, \citenamefont {Lee}, \citenamefont {Kim}, \citenamefont
  {Ryu}, \citenamefont {Lee}, \citenamefont {McMichael},\ and\ \citenamefont
  {Stiles}}]{Moon2013}%
  \BibitemOpen
  \bibfield  {author} {\bibinfo {author} {\bibfnamefont {J.-H.~H.}\
  \bibnamefont {Moon}}, \bibinfo {author} {\bibfnamefont {S.-M.~M.}\
  \bibnamefont {Seo}}, \bibinfo {author} {\bibfnamefont {K.-J.~J.}\
  \bibnamefont {Lee}}, \bibinfo {author} {\bibfnamefont {K.-W.~W.}\
  \bibnamefont {Kim}}, \bibinfo {author} {\bibfnamefont {J.}~\bibnamefont
  {Ryu}}, \bibinfo {author} {\bibfnamefont {H.-W.~W.}\ \bibnamefont {Lee}},
  \bibinfo {author} {\bibfnamefont {R.~D.}\ \bibnamefont {McMichael}}, \ and\
  \bibinfo {author} {\bibfnamefont {M.~D.}\ \bibnamefont {Stiles}},\ }\href
  {\doibase 10.1103/PhysRevB.88.184404} {\bibfield  {journal} {\bibinfo
  {journal} {Phys. Rev. B}\ }\textbf {\bibinfo {volume} {88}},\ \bibinfo
  {pages} {184404} (\bibinfo {year} {2013})}\BibitemShut {NoStop}%
\end{thebibliography}
%

\end{document}